# Optimizing GPU Cache Policies for MI Workloads

Johnathan Alsop*, Matthew D. Sinclair*[†], Srikant Bharadwaj*, Alexandru Dutu*, Anthony Gutierrez*, Onur Kayiran*, Michael LeBeane*,
Sooraj Puthoor*[†], Xianwei Zhang*, Tsung Tai Yeh*[‡], Bradford M. Beckmann*

*AMD Research, [†]University of Wisconsin – Madison, [‡]Purdue University

*Abstract*—In recent years, machine intelligence (MI) applications have emerged as a major driver for the computing industry. Optimizing these workloads is important but complicated. As memory demands grow and data movement overheads increasingly limit performance, determining the best GPU caching policy to use for a diverse range of MI workloads represents one important challenge. To study this, we evaluate 17 MI applications and characterize their behaviors using a range of GPU caching strategies. In our evaluations, we find that the choice of caching policy in GPU caches involves multiple performance trade-offs and interactions, and there is no one-size-fits-all GPU caching policy for MI workloads. Based on detailed simulation results, we motivate and evaluate a set of cache optimizations that consistently match the performance of the best static GPU caching policies.

## I. INTRODUCTION

In recent years, MI has emerged as an important driver for the computing industry. The initial catalyst for this rise in popularity was the discovery that MI could produce low error rates for image classification [1][2][3]. Subsequently, there has been a large amount of work optimizing hardware for MI, especially for Convolutional Neural Networks (CNNs) (e.g., [17]-[30]). Although these works have led to significant improvements in performance and energy efficiency of CNNs on modern multi-core CPUs, GPUs, and accelerators, it is challenging to analyze how future architectures will perform for these workloads. Here we focus on GPUs, as they are widely used for running MI workloads in numerous domains.

Although many MI systems use large discrete non-coherent GPUs instead of smaller cache coherent GPUs tightly coupled with the CPUs, the emerging trend is to unify the CPU-GPU memory system regardless of the GPU size [48]. Specifically, a single shared memory space between the CPU and GPU avoids the need for explicit data copies before and after every kernel launch. As a result, they are easier to program and can significantly reduce unnecessary data movement when GPU kernel launches are frequent, as can be the case with many MI workloads.

However, implementing efficient coherent caches between CPUs and GPUs remains a significant challenge. GPU workloads have very different memory demands from conventional CPU workloads. By concurrently executing hundreds to thousands of threads, GPUs can hide a large amount of memory latency, but they require a very high request bandwidth. This motivates a coherence strategy which prioritizes memory throughput and scalability, sometimes at the cost of cache reuse. In an effort to better understand the trade-offs of different caching strategies for MI workloads, we evaluate the performance of these applications with multiple levels of GPU caching enabled using the publicly available AMD gem5 APU simulator [5].

We find that there is no one-size-fits-all caching policy that offers the best performance to all MI workloads. Although caching can significantly improve performance by enabling local data reuse, in some cases the best caching policy is not the one that enables the most caching. The added blocking and contention introduced by caching can lead to harmful cache stalls and DRAM row locality disruption. In high throughput workloads that lack significant data reuse, the increased memory latency and decreased memory throughput caused by cache resource contention can make a simpler cache bypassing strategy more attractive.

Motivated by these results, we model and evaluate three microarchitectural optimizations which work together to mitigate these caching inefficiencies encountered by MI workloads. The first optimization avoids blocking for cache allocation, which reduces cache stalls. The second optimization applies a state-of-the-art CPU cache rinsing technique [58] to the last-level GPU cache to improve row buffer locality. Finally, we use a PC-based bypass prediction technique [54] to address remaining caching overheads while still caching accesses that can benefit from reuse. Collectively, these optimizations achieve the benefits of GPU cache reuse, while minimizing caching overheads for these important MI workloads.

In the remainder of this paper, we first cover relevant MI and GPU coherence background information in Sections II and III. Next, we introduce the system and applications we are using and discuss the changes that were necessary to run MI applications in Sections IV and V. In Section VI we use gem5 to evaluate MIOpen benchmarks running on a CPU-GPU cache hierarchy with a range of coherence policies. Based on this detailed data we motivate and evaluate a set of coherence optimizations in Section VII. We then further discuss related work (Section VIII), and finally conclude (Section IX).

## II. MI Background

Although there are many different MI methods, in this work we focus on deep neural networks (DNNs), which are some of the most commonly used MI workloads and are well-supported by MIOpen. CNNs and recurrent neural networks (RNNs) are two variants of DNNs, which are composed of multiple layers that apply linear and non-linear transformations (and other techniques like pooling) to iteratively reduce error and learn from the training data. Broadly speaking, DNNs use a combination of layers that are typically trained using backward propagation and stochastic gradient descent (SGD), and which generally access memory in a regular and dense manner. However, they can differ in the type and number of layers, as well as layer connectivity, which affects cache reuse potential, memory sensitivity, and bandwidth demand. It is crucial for memory system designers to understand the unique characteristics and potential performance bottlenecks of different DNNs and the layers within them, and to build hardware that can respond appropriately to dynamically changing memory properties.

### A. Neural Networks

Neural networks comprise multiple layers with various functionality, such as convolutional, activation, normalization, pooling, or fully connected layers. At the core of DNNs are **activation layers**. Activations such as Rectified Linear Unit (ReLU) are used to provide some non-linearity that helps to successfully train many networks. Because activation layers typically apply simple functions, they have low compute requirements. In addition, because an activation layer applies the activation function in an elementwise fashion, there is very low data reuse in these layers.

Although many neural network layers are sparsely connected convolutional layers, some are **fully connected**. Logically, such a layer connects every output neuron from the previous layer to every input neuron in consumer layer. Fully connected layers typically exhibit high reuse of both model weights and input elements, and they are compute-intensive.

### B. Convolutional Neural Networks

CNNs are sparsely connected neural networks with a form of connectivity [7]. **Convolutional layers** are at the heart of CNNs, and because in CNNs not every output unit interacts with every input unit, they do not need to learn as many parameters. This reduces the memory requirements of the model substantially. Since outputs will only share inputs with spatially local outputs, this limits the reuse potential relative to fully connected layers. Furthermore, computing the output for each layer requires less computation. Even so, in modern networks, convolutions typically dominate computational time.

Another innovation in network structure involves a process known as **pooling.** Pooling, also called down sampling, is a specialized hidden layer. A pooling layer is designed to replace the values from a small region with a single representative value and does not use neurons. *Max pooling*, a common method of pooling, retains the largest value in the region. Pooling reduces model *overfit*, or the tendency of a model to fit to the noise in the data instead of the actual signal. Data access in these layers is dense and regular, reuse is limited, and due to the unbalanced nature of the operation, load and store counts are unequal.

Normalization layers may also be used to facilitate faster convergence times for training by mitigating *covariate shift*. **Local response normalization (LRN)** and **batch normalization (BN)** are commonly used normalization layers. These techniques differ in which dimension they normalize (across batches or spatially within an input), but both have potential for input reuse with adjacent elements.

There are also specialized output layers. Many classifiers have output layers that use **softmax**. A softmax output layer normalizes the values in a neural network so that the entire vector sums to 1. The output of a classifier represents the probability that any element in the vector is the solution. The computation and data requirements are relatively minimal.

### C. Recurrent Neural Networks

RNNs have memory (represented by a hidden state vector), that allows them to capture information about what has happened previously. These networks contain loops that allow information to persist across multiple iterations. These loops can be unrolled such that each of the unrolled iterations passes information to the next iteration, with each cell performing one or more fully connected operations and activation functions. The hidden state is calculated by looking at the previous hidden state and the input at the current step, essentially operating as a fully connected layer to process the concatenated input. Unlike other neural networks, RNNs share parameters across all steps using a technique known as *weight sharing*, so the potential for reuse is even greater than for a fully connected layer. Even so, this is combined with activation kernels, which exhibit low reuse.

To help overcome the problem of *exploding gradients*, which affects conventional RNNs, researchers have introduced new types of RNNs such as **Long Short-Term Memory (LSTM)** and **Gated Recurrent Unit (GRU)** models. LSTM and GRU retain the same basic concept as a vanilla RNN but introduce a memory unit. The memory unit is a logical gate in the RNN that is designed to learn and retain long-term dependencies.

## III. CPU-GPU Cache Coherence Background

Tightly coupled CPU-GPU systems can greatly improve programmability and performance for heterogeneous workloads. Unlike discrete GPUs, which require explicit data transfer between the CPU and GPU memory space before and after every kernel launch, tightly coupled GPUs share a unified memory space and maintain coherence between the caches of each device via a shared system directory. Although this may add some complexity to the system design, it avoids unnecessary data transfer and latency for multi-kernel applications such as the RNNs and Composed Model (CM), which combines convolutional layers with pooling layers.

Table 1: Key simulated system parameters.

| GPU Parameters | |
|---|---|
| GPU Clock | 1600 MHz |
| # of CUs | 64 |
| # SIMD units per CU | 4 |
| Max # Wavefronts per SIMD unit | 10 |
| VRF/SRF per SIMD unit | 512/1600 |
| **CPU Parameters** | |
| CPU Clock | 4000 MHz |
| # CPUs | 2 |
| **Memory Hierarchy** | |
| GPU L1 D-cache per CU | 16 KB, 64B line, 16-way write-through |
| GPU L1 I-cache per 2 CUs | 32 KB, 64B line, 16-way |
| GPU L2 cache per 64 CUs | 4 MB, 64B line, 16-way write-through (write-back for R data) |
| Main Memory | HBM2, 16 GB, 16 channels, 16 banks/channel, 1000 MHz, 512GB/s |
| Approximate uncontested L1/L2/Memory latency | 50/125/225 cycles |

In order to understand trade-offs between different GPU cache policies in a CPU-GPU system, it is important to understand how coherence is implemented in such an environment. Unlike CPU codes, GPU workloads tend to be much more sensitive to memory throughput than memory latency, and GPU caching policies are therefore designed to be simpler and more scalable than conventional CPU protocols. Rather than requesting and tracking read and write permissions for accesses, GPU caches simply write-through written data and self-invalidate read data at synchronization points (i.e., kernel boundaries) [52][53]. This avoids the overheads of reader/writer tracking and writer-initiated invalidation, enabling caches to scale to the higher throughput demands of GPU workloads. In the system we study (Section V.B), a shared L2 is then used to filter and coalesce requests before they interface with a more conventional (and complex) CPU coherence fabric.

However, as we will show, even this simple strategy can miss some performance opportunities for some GPU MI workloads. Depending on the application, bypassing the cache for some or all data accesses can lead to better performance by avoiding these overheads.

In this work, we explore the costs and benefits of caching in GPU MI workloads by simulating three caching policies that differ in how loads and stores are handled in GPU caches:

- Uncached: Loads and stores bypass all GPU caches.
- CacheR: Loads are cached in L1 and L2, but stores bypass all GPU caches.
- CacheRW: Loads are cached in L1 and L2, stores bypass L1 and are combined in L2.

When load caching is disabled, read requests to the same cache line may be coalesced while the original bypass request is pending, but on a response the data is forwarded without being inserted in the cache. When load caching is enabled (CacheR, CacheRW), the GPU L1 and L2 caches always self-invalidate valid data at synchronization points (e.g., kernel boundaries) [53]. When store caching is enabled (CacheRW), stores still bypass the L1 but they may be coalesced at the L2 until a flush of all L2 dirty data is triggered at a system-scope synchronization point, at which time they are written back to memory [61]. The performance effects of each policy, as well as optimizations that target some of the inefficiencies discovered, are discussed in Section 5.

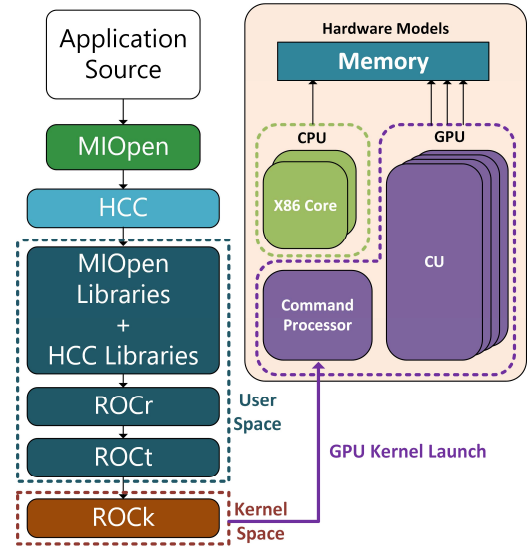

Figure 1: ROCm gem5 compilation flow.

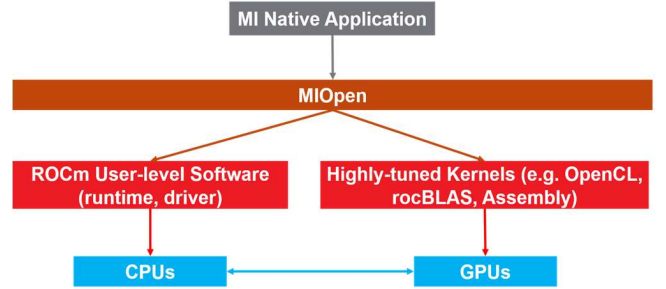

Figure 2: MI application flow.

## IV. EMULATION PROCESS

In this section we describe our process for running MI workloads on a simulated APU in gem5. Figure 1 shows how the gem5 simulator runs MIOpen applications, including the HCC [14][15] and HIP [16] libraries, on an unmodified version of the ROCm user-level stack. Since prior work discusses the compilation flow in detail [5][6], we instead focus on the changes to this flow that are needed to simulate the MIOpen framework and the intermediary libraries like rocBLAS and MIOpenGEMM from Figure 2. The gem5 simulator models a system with multiple CPUs and a GPU with multiple compute units (CUs). The CPU and GPU are coherently coupled together, which is an emerging trend in CPU-GPU systems [48]. Thus, the CPU and GPU share a single unified cache coherent address space and do not require explicit copies.

The gem5 simulator can simulate the entire system in full system (FS) mode, including devices and an operating system, or emulate the system calls in syscall emulation (SE) mode. SE mode only simulates user-space execution and provides system services (e.g., system calls) in the simulator instead of executing OS kernel-space code. In this work, we use SE mode because GPU kernels do not make system calls; instead they rely on rich user-space libraries like the ROC runtime (ROCr) to provide many system services and to do device configuration and setup. Furthermore, we use SE mode while executing the off-the-shelf ROCm stack, which does the bulk of the system work. Therefore, our methodology only emulates the lowest levels of the software stack and preserves the fidelity of all user-level software components. As a result, the only portion of the ROCm software stack that must be emulated is the ROCm Linux kernel driver.

One of the main changes required for simulating MI applications in gem5 was extending MIOpen to support APUs. Currently MIOpen mainly targets discrete GPUs. Thus, we modified the libraries in Figure 2 to generate code for APUs. Specifically, we modified the applications, HIP, and MIOpen to remove the device copies wherever possible. We also rebuilt HIP to perform all memory management on the host instead of the device. This was necessary because part of the default rocBLAS library for discrete GPUs is hardcoded to use device copies. As a result, both the CPU and GPU almost always use the same copy of the data. Using open source libraries was a key enabler in overcoming this challenge, because we could recompile the libraries after making the necessary changes.

In addition to the changes required to simulate the MIOpen applications in gem5, we also made additional changes to reduce simulation execution time. First, MIOpen uses clang-ocl to perform online compilation of kernels the first time it executes an application. As part of this process, MIOpen caches each kernel binary, to avoid recompiling on a subsequent use of the kernel. Since online compilation is computationally intensive and not part of the application's region of interest, we bypass online kernel compilation in gem5 by running the applications on a real AMD APU beforehand to obtain MIOpen's cached kernel binaries. In some cases, the need for additional kernels when we added new applications was lessened because other MI applications used the same MIOpen kernels.

Second, for every application that uses GEMM kernels, MIOpenGEMM will create a database of GEMM kernels and then select the kernel that best matches the application's matrix size(s). To avoid the overhead of simulating this process, we added logic to bypass the on-line kernel database creation for the most popular GEMM kernels. Thus, MIOpenGEMM only creates its database when it encounters unpopular GEMM kernels (e.g., a GEMM size that no prior application had used).

Finally, DNNMark and MIOpen-benchmark also perform a sweep of every possible kernel that could be used for each layer of the neural network. Afterwards, the benchmarks measure the total execution time of the fastest option for each layer. This makes sense on real GPUs, where kernels execute relatively quickly, and the primary goal is performance benchmarking. However, this greatly increases simulation time and our main goal is to evaluate relative performance differences of potential hardware features. Thus, like other solutions, we added bypass logic in DNNMark and MIOpen-benchmark that preselects the fastest kernel for a given layer based on the execution time on a real AMD APU.

## V. METHODOLOGY

### A. The gem5 Simulator

To analyze how MI workloads perform on future architectures and the benefits of co-designed hardware-software solutions, we leverage the gem5 simulator [4][5]. The gem5 simulator is a natural choice for this work because it models both the CPU and GPU with high fidelity, including multi-threaded synchronization and cache coherence. Other tools also attempt to project MI performance, but they either have not been released [42][44][45][46], exclusively focus on modeling the GPU kernel execution [47], or have been released but only optimize the neural network before execution (e.g., XLA [32] and ONNX [33]). Although some of these approaches could eventually be incorporated into gem5, in this work we instead focus on executing both the CPU and GPU parts of open-sourced MI applications. More recently, GPGPGU-Sim [60] and Multi2Sim [35][59] have been updated to support MI workloads; these simulators could also be used for this study and we expect they would provide similar results.

Prior work has shown how to use AMD's ROCm ecosystem to simulate HCC and HIP applications in gem5 with high fidelity compared to an AMD APU [5][6]. In this work, we build from the existing ROCm support to simulate MI applications that use the MIOpen library [36]. Although there are several widely used MI libraries, MIOpen is one of the few open-source libraries, which allowed us to easily change it to work with cache coherent APUs. We discuss other GPU simulators and modeling techniques in Section VIII. We extend prior work on running AMD APUs in gem5 to execute MI applications from a wide variety of suites, including DNNMark [8], DeepBench [9][10], and MIOpen-benchmark [11]. These suites cover a wide range of MI uses, including CNNs and RNNs.

### B. System Configuration

Table 1 lists the key system parameters we simulate in gem5.[1] Figure 3 shows the conceptual system design, which includes a 64-CU GPU with two levels of cache [37]. Our simulated GPU CU pipeline is based on AMD's GCN architecture [49] and uses the GCN3 ISA [41]. We model single-cycle instruction issue with 64-wide wavefronts and the model uses 64-wide SIMDs. Our detailed model also models the delay to process packets and its memory accesses. Similarly, the host CPU model uses gem5's detailed out-of-order, superscalar, pipelined x86-64 processors with SMT support.

---
[1] Without loss of generality, we use the AMD GPU terminology. The NVIDIA CUDA equivalents for these terms are SM (CU), threads (work items), warps (wavefronts), and thread blocks (work groups).

Table 2: Studied MI workloads.

| Application | Input | Unique Kernels/ Total Kernels | GPU Footprint |
|---|---|---|---|
| Backward Activation (BwAct) [8] | Batch size 100 | 1/1 | 2.4 GB |
| Backward Batch Normalization (BwBN) [8] | Batch size 512 | 1/1 | 5.88 MB |
| Backward Pool (BwPool) [8] | Batch size 256 | 1/1 | 252 MB |
| Backward Softmax (BwSoft) [8] | Batch size 512 | 1/1 | 0.02 MB |
| Composed Model (CM) [8] | Batch size 64 | 4/130 | 12.1 MB |
| Forward Activation (FwAct) [8] | Batch size 100 | 1/1 | 1.6 GB |
| Forward Batch Normalization (FwBN) [8] | Batch size 256 | 1/1 | 42 MB |
| Forward Fully Connected (FwFc) [8] | Batch size 512 | 1/1 | 148.2 MB |
| Forward LRN (FwLRN) [8] | Batch size 100 | 1/1 | 2.4 GB |
| Forward Pool (FwPool) [8] | Batch size 256 | 1/1 | 480 MB |
| Forward Softmax (FwSoft) [8] | Batch size 512 | 1/1 | 0.01 MB |
| SGEMM [9][10] | 4Kx128x4K | 1/1 | 68 MB |
| DGEMM [9][10] | 4Kx128x4K | 1/1 | 132MB |
| RNN Forward (FwLSTM/GRU) [9][10] [11] | Batch size 1, sequence length 16, hidden layer 128, LSTM/GRU | 4/150 | 0.38 MB |
| RNN Forward Backward (FwBwLSTM/GRU) [9][10] [11] | Batch size 1, sequence length 16, hidden layer 128, LSTM/GRU | 6/363 | 0.48 MB |

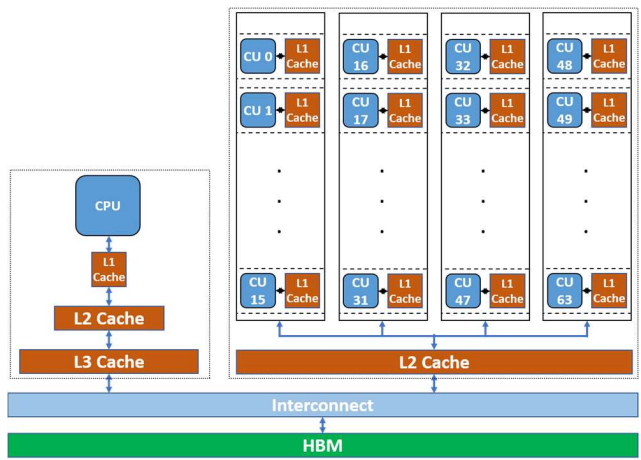

Figure 3: Overall simulated system.

## C. Applications

Table 2 shows the seventeen MI benchmarks that we studied. These MI benchmarks come from several popular MI suites: DNNMark [8], DeepBench [9][10], and MIOpen-benchmark [11]. We selected these benchmarks because they cover many different types of CNN and RNN layers and full NNs. Most of these benchmarks are single CNN layers, which make up the larger CNNs used in many DNNs. However, we also include several benchmarks such as the RNNs, and Composed Model that are larger MI applications. The input sizes for these workloads were selected based on the largest input sizes for these workloads that we could simulate in a reasonable amount of time (up to 5 days of simulation time), and we use the same inputs as the benchmark suites (randomly initialized values). All of these benchmarks call MIOpen directly.

Many of the workloads in Table 2 execute a single GPU kernel. These benchmarks, including most of the DNNMark benchmarks, often run a single CNN layer or operation. Studying these benchmarks is useful for examining the correctness of the simulator and studying the microarchitectural and memory behavior of specific layers within larger MI workloads. Moreover, they represent the building blocks that are used for the larger networks such as Composed Model and the RNN workloads.

The DeepBench RNN training and inference workloads in Table 2 are highly configurable, and can model many different sequence lengths, hidden layer sizes, and batch sizes. As hidden layer size, sequence length, and batch size increase, the number of kernels and GPU footprint also increase. Thus, these workloads are useful for examining the behavior of a variety of different RNN training and inference sizes. In this paper we present LSTM and GRU data since they are the most widely used variants and show results for an input that is representative of RNNs used in real English-Vietnamese speech translation RNNs [31].

## VI. CACHING CHARACTERIZATION

We begin by characterizing benchmark properties that are relatively independent of cache policy. To the best of our knowledge, this is the first characterization of MI CPU-GPU workloads in a coherent shared memory environment. Figure 4 shows the giga vector operations per second (GVOPS) for each workload, and Figure 5 shows the giga GPU memory requests per second (GMR/s) issued to the memory system for each workload (the CacheR policy is used for both). This data provides some insight into which workloads are compute bound and which are memory bound. In state-of-the-art MI kernels, the tiling pattern, work item/work group parallelism, and scratchpad memory usage can vary even for a given kernel based on what the framework determines is optimal for the target platform, making it difficult to generalize about the memory sensitivity of any class of MI tasks. However, as a general rule, workloads with low compute bandwidth and high memory request bandwidth are more likely to be sensitive to caching policy than workloads with low memory request bandwidth and high compute bandwidth.

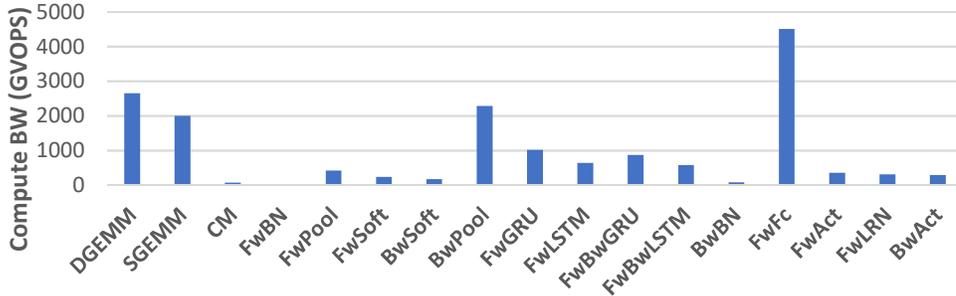

Figure 4: Giga vector ops per second with CacheR policy.

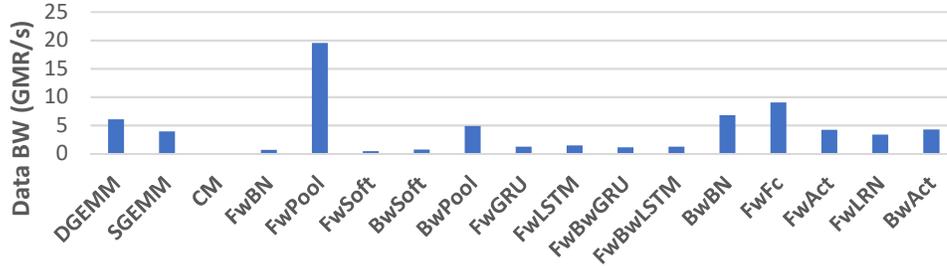

Figure 5: Giga memory requests per second with CacheR policy.

## A. Caching Policy Comparison

Figure 6 shows the execution time of each caching policy described in Section III for all applications, normalized to Uncached. Figure 7 shows the number of memory accesses that reach the DRAM controller, also normalized to Uncached. Overall, our results show that the best performing caching policy varies widely depending on the available cache reuse and memory sensitivity of the workload. Workloads can be grouped into three categories based on how they are affected by caching:

1. **Memory Insensitive**: Cache policy does not significantly affect overall execution time (<5% change) for CM, SGEMM, and DGEMM because the workload is compute bound or the potential for reuse is low.
2. **Reuse Sensitive**: Enabling caching consistently improves cache reuse and performance for FwBN, FwPool, FwSoft, BwSoft, BwPool, FwGRU, FwLSTM, FwBwGRU, FwBwLSTM, BwBN, and FwFc.
3. **Throughput Sensitive**: Enabling caching consistently hurts performance for FwAct, FwLRN, and BwAct due to a lack of cache reuse and high throughput demand for data.

## B. Caching Benefits: Reuse

The main benefit of caching is that it enables cache reuse and therefore lower latency and higher bandwidth access to data. Figure 7 shows the total number of GPU memory accesses issued to DRAM, normalized to Uncached. The reduction in this value represents the proportion of memory accesses that hit in the cache and gives a measure of read and write reuse potential. For most applications (excluding throughput sensitive applications), enabling read or write caching increases the proportion of accesses that hit in the caches.

The amount of added reuse enabled by caching is dependent on the amount of reuse in the algorithm and the amount of reuse possible with caching disabled. When caching is disabled, local reuse can still be exploited in two ways: 1) if accesses to reused data are from work items in the same work group, a kernel may use local data store (LDS) memory to load data once from memory then reused multiple times by threads in the work group, and 2) if accesses to the same data arrive close together in the cache, they can be coalesced until the original bypass request completes. With caching enabled, reuse between any threads over any period of time can be exploited in the caches.

As expected, the throughput sensitive workloads, which are the activation and normalization layers with no potential for reuse, experience no performance benefits or memory demand reduction from enabling caching.

In contrast, enabling caching improves reuse for the cache insensitive workloads, but this does not result in performance gains. For DGEMM and SGEMM, caching reads reduces memory demand by 74% and 84%, respectively, but these workloads are ultimately limited by compute throughput. Read and write caching improve reuse by 69% for CM, but performance is unaffected due to CM's exceptionally low memory demand.

The remaining reuse sensitive workloads benefit from read and/or write caching to various degrees. Layers with limited connectivity, such as the pooling, convolution, and some normalization layers show limited benefit in large part because reuse is primarily between nearby work items and can be exploited even when caching is disabled. However, for workloads with higher connectivity, where reuse is possible between distant work items (e.g., FwFC, FwBN, FwBwGRU, and FwBwLSTM), we find that read caching can reduce memory demand by up to 93%. When the accesses that experience reuse are critical to performance, this can also result in performance gains, reducing execution time by up to 29%. In addition, write caching can further reduce memory demand by up to 71% and execution time by up to 32% for reuse sensitive workloads which exhibit high potential for write coalescing at L2 such as BwPool and BwBN.

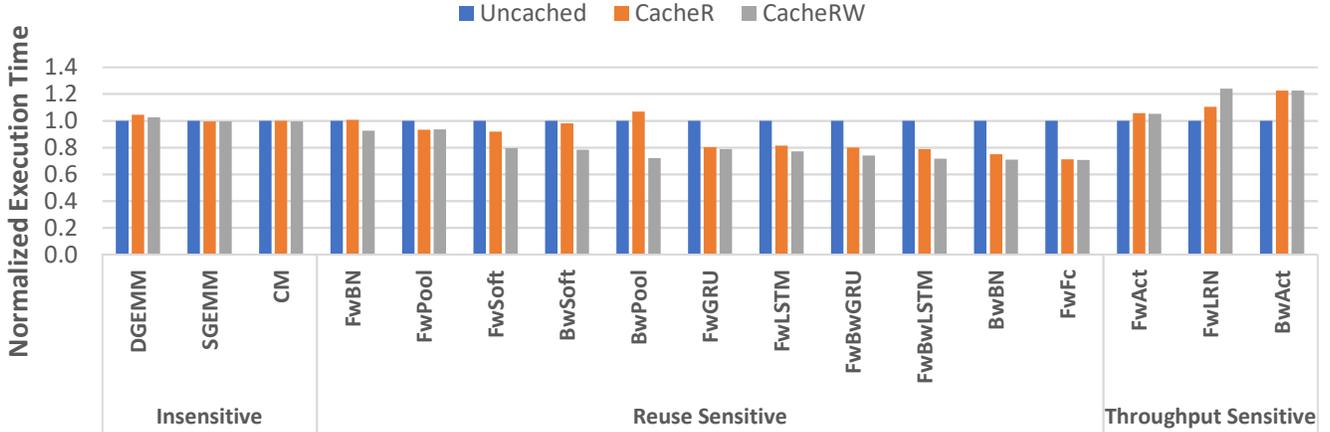

Figure 6: Execution time for all applications using each cache policy. Normalized to Uncached.

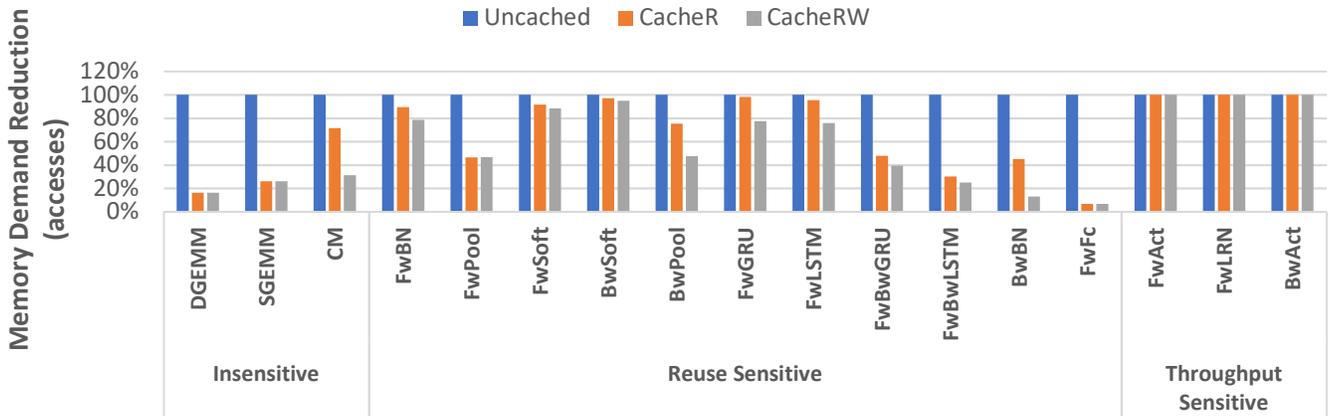

Figure 7: Number of GPU memory requests which reach DRAM, normalized to Uncached.

### C. Caching Overheads

Improved cache hit rates do not tell the whole story. For throughput-sensitive workloads, the overheads of caching may outweigh the benefits. We find that coherence overheads manifest themselves primarily as 1) cache stalls due to added contention for cache resources, and 2) reduced DRAM row locality for requests that have been delayed in the caches.

*1) Cache Stalls*

We define a cache stall as any cycle in which a ready cache request is blocked from querying a cache at any level. When caching is enabled, cache operations can increase cache stalls in multiple ways. Cached requests require allocation on a miss, and this may cause stalls if all lines in the set are in a busy state (e.g., waiting for a pending load). In addition, coherence operations can add contention for shared resources such as tag arrays (e.g., due to failed cache allocation).

Figure 8 plots of cache stall counts on a logarithmic scale normalized to the number of GPU L1 requests. High cache stall counts lead to worse execution time for FwAct and BwAct, when read caching is enabled. FwPool also experiences high cache stalls, although any negative effect here is offset by the added reuse achieved through caching.

*2) DRAM Row Locality*

Enabling read or write caching adds variability to memory access times through cache stalls described in Section 1)VI.C.1) or by delaying stores at the L2 so they can be coalesced (CacheRW). For programs with highly regular access patterns and limited reuse potential, this added variability can negatively impact DRAM row locality, which in turn limits achievable DRAM bandwidth and increases memory latency. Figure 9 shows how different cache policies affect the row hit ratio for DRAM loads and stores. MI applications tend to have regular access patterns, and as a result enabling caching can interfere with this regularity and hurt DRAM row hit rates[2]. In particular, FwPool, FwAct, FwLRN, and BwAct suffer from this effect. Although this effect in FwPool is outweighed by the benefits of cache reuse, in the throughput sensitive workloads it contributes to a performance degradation for caching configurations.

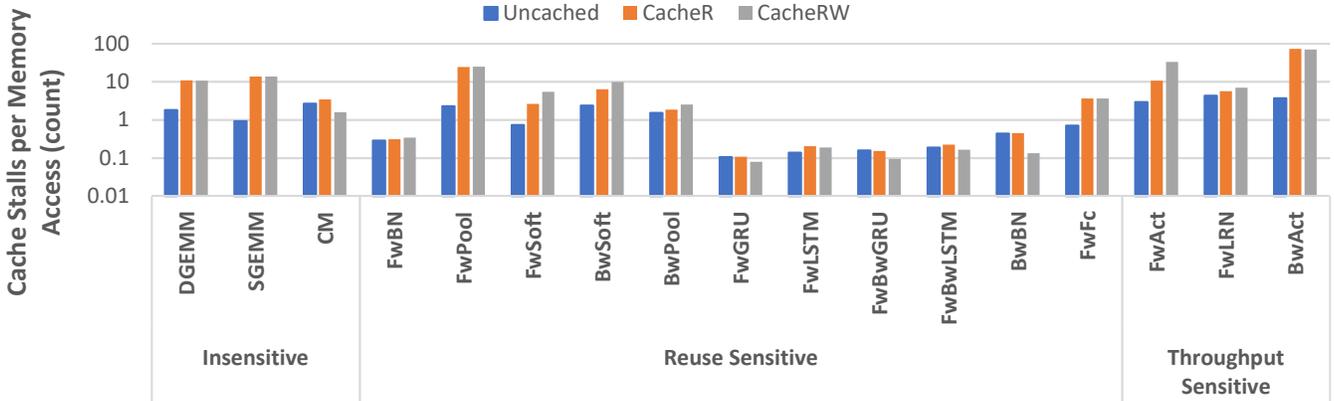

Figure 8: Cache stall count (log scale) for all applications normalized to the total GPU memory request count.

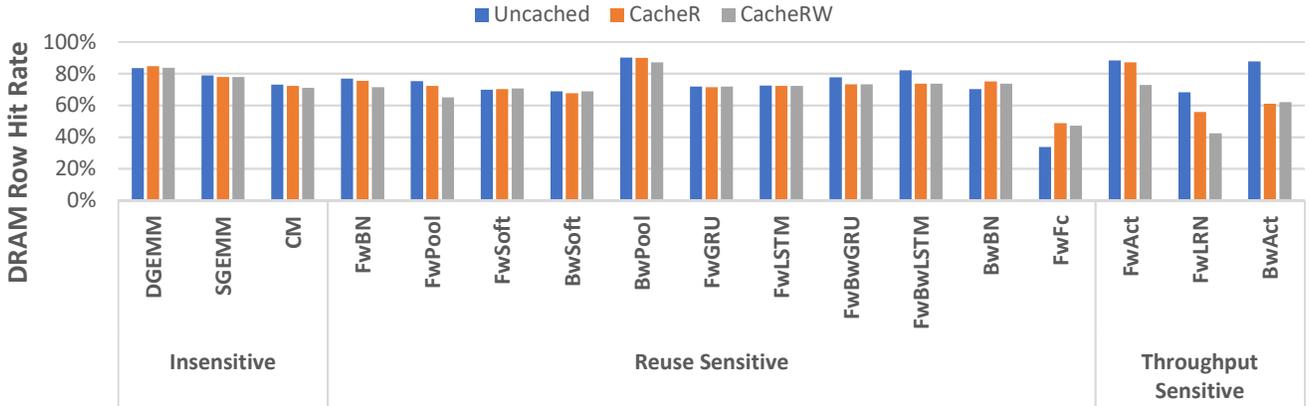

Figure 9: DRAM row buffer hit ratios for all applications.

## VII. CACHING OPTIMIZATIONS FOR MI APPLICATIONS

Motivated by the caching overheads we observe in GPU MI workloads, we next describe three potential architectural optimizations and evaluate their effect on performance. All are applied to the most aggressive caching policy, CacheRW, and are compared against the best and worst performing static configurations as measured in Figure 6. Figure 10 reports the normalized execution time for these optimizations, Figure 11 plots the relative number of DRAM accesses for each configuration, Figure 12 shows cache stall counts per GPU memory request plotted on a logarithmic scale, and Figure 13 reports DRAM row hit rates.

### 1) Allocation Bypass

We begin by attempting to address the overhead of cache stalls due to blocked cache allocation operations. When caching is enabled and memory accesses require cache allocation, it may be necessary to stall the incoming request if all lines in the target set are occupied by a pending load or store until the blocking request completes. However, as we have seen this can limit bandwidth and disrupt DRAM row locality. To address this, we adapt our caching policies by converting cached requests to bypass requests whenever allocation would require blocking. This allocation bypass optimization is plotted as CacheRW-AB in Figure 10-Figure 13.

Although the non-blocking caching optimization reduces cache stalls per request significantly, it has a minimal effect on overall performance for most applications. This can be explained by the fact that allocation bypassing does little to reduce the added congestion overhead, and in some cases adds to it by eliminating a throttling effect from the L1 cache level (note the 7% higher execution time for FwPool). The main exception is FwLRN, which sees significant benefits due to improved DRAM row hit ratios. FwLRN is most affected by DRAM row locality disruption due to blocking allocation. Avoiding allocation blocking eliminates the predominant source of caching overhead, although a slight performance degradation remains from disruption due to coalesced delayed L2 store requests.

---

[2] The exceptions are BwBN and FwFC, which see higher row hit rates with caching enabled because caching filters interleaved repeated accesses such that primarily the highly regular compulsory misses make it to DRAM.

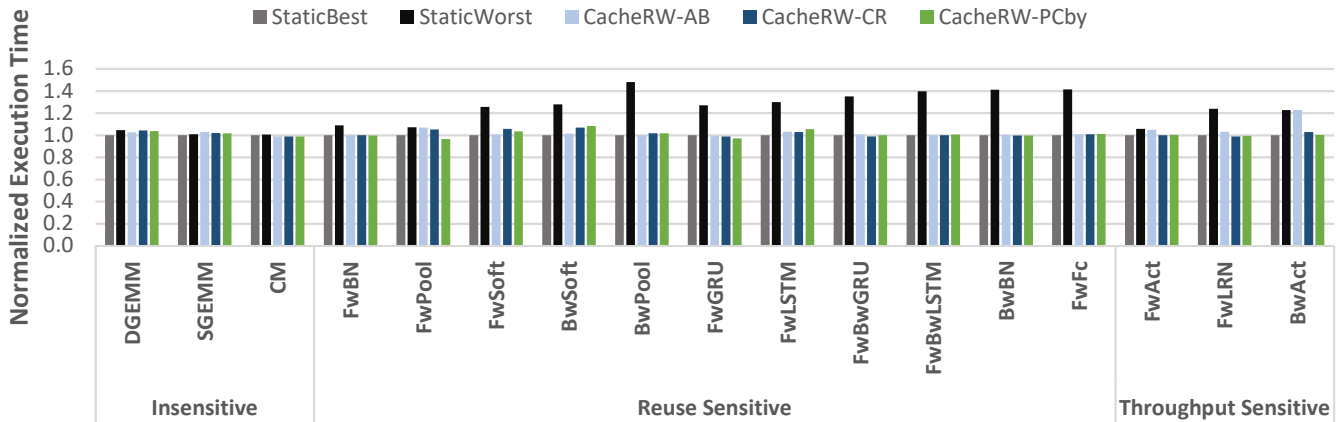

Figure 10: Execution time of best and worst static cache policy (from Figure 6) compared with allocation bypassing (CacheRW-AB), cache rinsing (CacheRW-CR), and PC-based bypassing (CacheRW-PCby) optimizations added. Normalized to best static configuration.

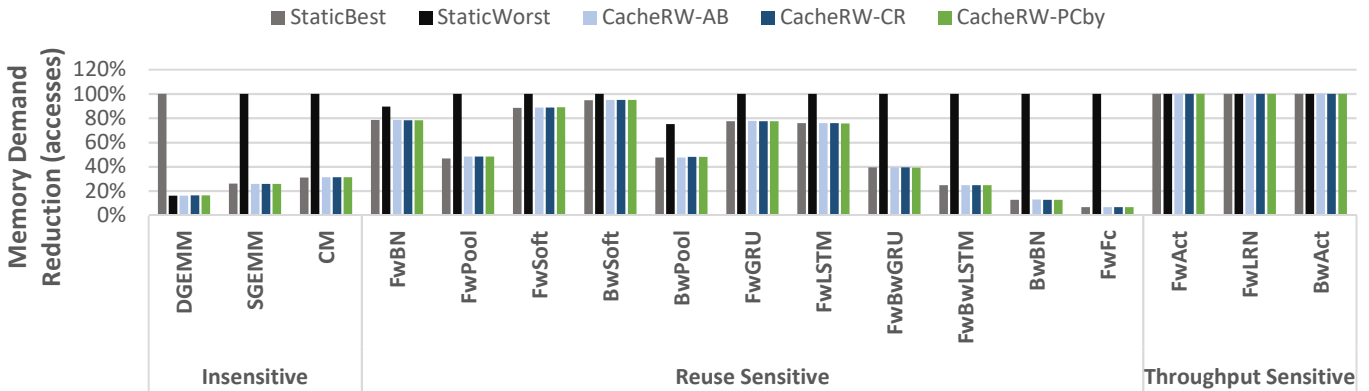

Figure 11: Number of GPU memory requests which reach DRAM for static best and static worst cache policy (from) compared with allocation bypassing (CacheRW-AB), cache rinsing (CacheRW-CR), and PC-based bypassing (CacheRW-PCby) optimizations added. Normalized to Uncached.

### B. Row Locality-Aware Cache Rinsing

Although allocation bypass avoids row locality disruption caused by blocking allocation operations, it does not avoid disruption due to L2 write coalescing. To address this, we next add a row locality-aware cache rinsing scheme based on a method originally proposed for CPUs by Seshadri *et al.* [58]. This technique adds a dirty block index to the GPU L2 that tracks dirty blocks in each DRAM row. Whenever a dirty block is evicted, a writeback of all other dirty blocks in that row is triggered.

We add this cache rinsing optimization on top of the allocation bypassing optimization, denoted CacheRW-CR in Figure 10-Figure 13. Cache rinsing counteracts the DRAM row locality overhead of caching for affected (mainly throughput sensitive) workloads, offering DRAM row hit rates that are even higher than those of the best static configuration. As Figure 10 shows, caching overheads in BwAct and FwAct are reduced as a result of this technique.

### C. PC-Based L2 Bypassing

We next attempt to address any remaining performance overheads due to caching by predicting whether caching will be beneficial (i.e., whether cache reuse is likely) then dynamically choosing to use cached requests and incur the resulting overheads only when that is the case. Past work has explored this concept for adaptive load bypassing at the L1, proposing a PC-based reuse predictor to avoid cache pollution and more effectively use limited cache space [54]; we apply the same PC-based technique instead to the L2 for both loads and stores for the purpose of avoiding congestion overheads when reuse is unlikely.

PC-based L2 bypassing is applied on top of the allocation bypassing and cache rinsing optimizations and denoted as CacheRW-PCby in Figure 10-Figure 13. Overall, it is effective at predicting reuse for MI workloads. For nearly all workloads, the combination of allocation bypassing, cache rinsing, and PC-based bypassing matches or exceeds the performance of the best static cache configuration by selectively incurring cache overheads when they are expected to be beneficial. For example, PC-based bypassing is able to overcome the overhead introduced to FwPool from allocation bypassing.

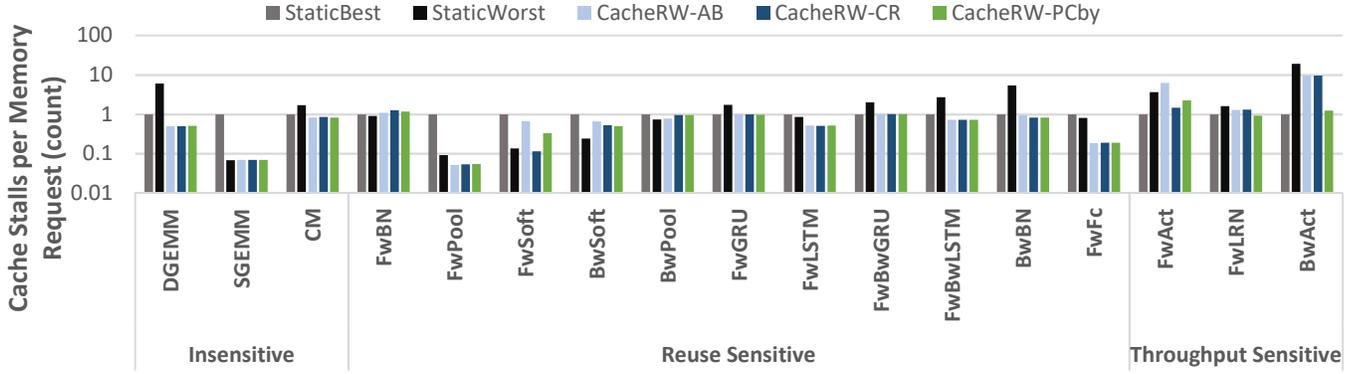

Figure 12: Cache stalls per memory request for best and worst static cache policy (from Figure 6) compared with allocation bypassing (CacheRW-AB), cache rinsing (CacheRW-CR), and PC-based bypassing (CacheRW-PCby) optimizations added. Plotted on logarithmic scale.

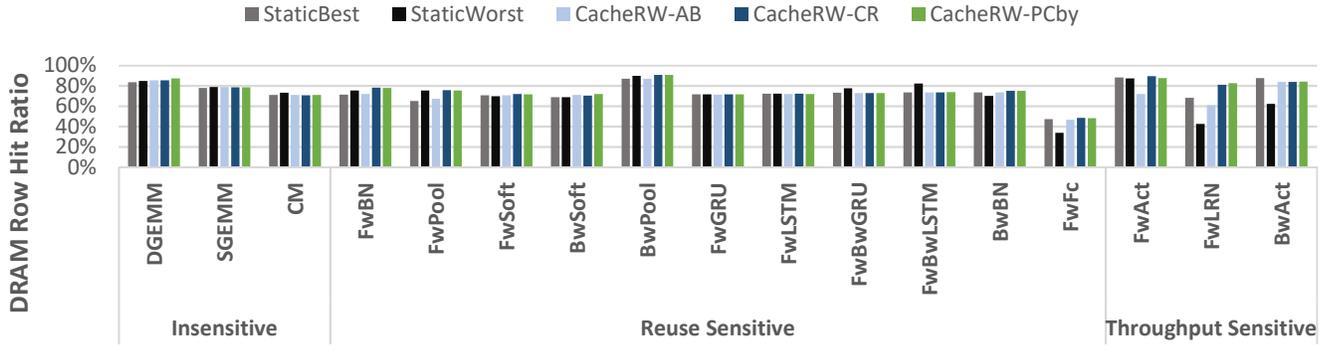

Figure 13: DRAM row hit ratio request for best and worst static cache policy (from Figure 6) compared with allocation bypassing (CacheRW-AB), cache rinsing (CacheRW-CR), and PC-based bypassing (CacheRW-PCby) optimizations added.

## VIII. RELATED WORK

There have been multiple prior efforts to enable efficient caching and coherence in GPUs [53][61][62][63][64][65][66]. In general, these aim to maximize cache reuse, often by avoiding or mitigating the cost of bulk flush and invalidation actions that are required for synchronization in GPU caches. In contrast, the techniques described in this work target MI workloads where reuse may be fundamentally limited by low locality, and where even the overheads of a simple caching mechanism can degrade performance. These prior efforts therefore miss out on potential performance improvements by focusing on increasing reuse rather than reducing overhead.

Past work has also proposed techniques for adapting GPU caching and coherence strategies to the access patterns of executing workloads. Whether through adaptive cache bypassing [54][69][70], locality-aware rinsing [55][56][57][58] (which to our knowledge has previously only been applied to CPU systems), flexible coherence request types [67], or cross-layer coordination of scheduling and memory management [78], these techniques can greatly improve cache efficiency by matching caching policies to GPU workload demands. In contrast, the primary contribution of this work is to characterize the sources of cache inefficiency in GPU MI workloads and to show how multiple techniques described above can be combined in a targeted and cooperative manner to address the specific caching overheads in this important domain.

To the best of our knowledge, this is the first work that characterizes MI workloads on a cycle-level, publicly available simulator like gem5. Prior work has simulated MI workloads on in-house simulators [42][44][45][46], but few details are available. In addition to using in-house simulators, SCNN and CDMA use analytical models, like TimeLoop, to analyze NNs [42][43]. CDMA explicitly mentions the lack of publicly available MI simulators as one reason for using analytical models, which enhances the importance and necessity of this work. More recently, GPGPGU-Sim [60] and Multi2Sim [35][59] have been updated to support MI workloads. Although these tools could be used to perform similar studies, GPGPU-Sim focuses on discrete GPUs, not tightly coupled ones like gem5 models (GPGPU-Sim's recent updates could be integrated into gem5-gpu [34] to allow such a study). Moreover, recent work has shown that simulating at a higher level, like GPGPU-Sim, loses important architectural details and may lead to incorrect conclusions [5]. Thus, when combined with the importance of being able to recompile the libraries to run APU-compliant code, we believe gem5 and MIOpen represent the best combination between simulator and MI library.

There has also been prior work on analyzing MI workloads at a higher level [38][39][40] than the lower level support we added to gem5. However, these efforts are unable to obtain the same level of detailed analysis (e.g., coherence traffic and stall cycles) that we can obtain with gem5. Projects like XLA [32], ONNX [33], and DeepCPU [50] optimize the entire neural network before execution. Although these projects also optimize MI workloads, we view these works as complementary to ours, and potentially something that could be added on top of our framework.

## IX. Conclusions and Future Work

In this work, we demonstrate that the gem5 APU simulator can execute CPU+GPU MI applications using MIOpen. Using this tool, we characterize the performance effects of a variety of GPU caching policies on these workloads. Overall, we found that caching reads and writes has mixed behavior for MI workloads. For some workloads, it improves performance by up to 29% through increased cache reuse, but for others it degrades performance by up to 24% by incurring cache stalls and disrupting DRAM row locality. Based on these trade-offs, we motivate and evaluate a set of adaptive cache optimizations with gem5. For most applications, these optimizations allow us to leverage the benefits of caching when it is helpful while avoiding performance overheads when caching hurts.

These results demonstrate that, although the MI applications studied all exhibit regular and dense memory access patterns, there is no statically ideal caching policy for GPU MI workloads. Thus, being able to selectively avoid coherence overheads where possible is important to performance. In the future, GPU coherence overheads will likely only grow in importance. Future heterogeneous systems will likely require even deeper cache hierarchies and non-uniform memory access interfaces, while MI workloads may exhibit even higher throughput demands and more frequent synchronization. As bulk coherence operations become more complex, expensive, and frequent, it is critical to be able to understand these trade-offs and address them with smart and adaptive cache policies.

More broadly this paper demonstrates the robustness of the AMD gem5 APU simulator and the benefit of open-source software stacks. In the important MI domain, researchers need the tools to evaluate and enhance the entire system solution space. Using gem5 to explore the space of caching strategies, we show how researchers can now rapidly prototype new hardware features running production-quality software and examine how they behave in tightly coupled CPU-GPU systems.

## X. Acknowledgements

The authors would like to thank Gabe Loh for his valuable feedback on this work. AMD, the AMD Arrow logo, Radeon, and combinations thereof are trademarks of Advanced Micro Devices, Inc. Other product names used in this publication are for identification purposes only and may be trademarks of their respective companies.

## References

[1] Olga Russakovsky, Jia Deng, Hao Su, Jonathan Krause, Sanjeev Satheesh, Sean Ma, Zhiheng Huang, Andrej Karpathy, Aditya Khosla, Michael Bernstein, Alexander C. Berg, and Li Fei-Fei. 2015. ImageNet Large Scale Visual Recognition Challenge. In *Int. J. Comput. Vision* 115, 3 (December 2015), pp. 211-252.

[2] Alex Krizhevsky, Ilya Sutskever, and Geoffrey E. Hinton. "ImageNet classification with deep convolutional neural networks." *NIPS*, 2012.

[3] Kaiming He, Xiangyu Zhang, Shaoqing Ren, and Jian Sun. "Delving deep into rectifiers: Surpassing human-level performance on ImageNet classification." In *Proceedings of the IEEE International Conference on Computer Vision*. 2015.

[4] Nathan Binkert, Bradford Beckmann, Gabriel Black, Steven K. Reinhardt, Ali Saidi, Arkaprava Basu, Joel Hestness, Derek R. Hower, Tushar Krishna, Somayeh Sardashti, Rathijit Sen, Korey Sewell, Muhammad Shoaib, Nilay Vaish, Mark D. Hill, and David A. Wood. The gem5 Simulator. May 2011, In *ACM SIGARCH Computer Architecture News*.

[5] Anthony Gutierrez, Bradford M. Beckmann, Alexandru Dutu, Joseph Gross, John Kalamatianos, Onur Kayiran, Michael LeBeane, Matthew Poremba, Brandon Potter, Sooraj Puthoor, Matthew D. Sinclair, Mark Wyse, Jieming Yin, Xianwei Zhang, Akshay Jain, Timothy G. Rogers. Lost in Abstraction: Pitfalls of Analyzing GPUs at the Intermediate Language Level. In *Proceedings of the 24th IEEE International Symposium on High-Performance Computer Architecture (HPCA)*, February 2018.

[6] Anthony Gutierrez, Bradford M. Beckmann, Sooraj Puthoor, Matthew D. Sinclair, Tuan Ta, and Xianwei Zhang. AMD gem5 APU simulator: Modeling GPUs Using the Machine ISA. *Tutorial at International Symposium on Computer Architecture*, June 2018.

[7] Yann LeCun. "Generalization and network design strategies." Connectionism in perspective (1989): pp. 143-155.

[8] Shi Dong and David Kaeli. DNNMark: A Deep Neural Network Benchmark Suite for GPUs. In *Proceedings of the General Purpose GPUs* (GPGPU-10). 2017.

[9] Sharan Narang. DeepBench. https://svail.github.io/DeepBench/. September 2016.

[10] Sharan Narang and Greg Diamos. An update to DeepBench with a focus on deep learning inference. https://svail.github.io/DeepBench-update/. June 2017.

[11] Pat Flick. MIOpen-benchmarks. https://github.com/patflick/miopen-benchmark. October 2017.

[12] Andrej Karpathy. The Unreasonable Effectiveness of Recurrent Neural Networks. http://karpathy.github.io/2015/05/21/rnn-effectiveness/. May 2015.

[13] Christopher Olah. Understanding LSTM Networks. http://colah.github.io/posts/2015-08-Understanding-LSTMs/. August 2015.

[14] AMD. HCC: An open source C++ compiler for heterogeneous devices. https://github.com/RadeonOpenCompute/hcc.

[15] Ben Sander, Greg Stoner, Siu-chi Chan, Wen-Heng Chung, Robin Maffeo. HCC: A C++ Compiler for Heterogenous Computing. HSA Foundation, Tech Report (2015).

[16] HIP: Heterogeneous-computing Interface for Portability. https://github.com/ROCm-Developer-Tools/HIP/.

[17] Amir Yazdanbakhsh, Kambiz Samadi, Nam Sung Kim, and Hadi Esmaeilzadeh. GANAX: A Unified MIMD-SIMD Acceleration for Generative Adversarial Networks. *Proceedings of the 45th Annual Symposium on Computer Architecture (ISCA)*, June 2018.

[18] Animesh Jain, Amar Phanishayee, Jason Mars, Lingjia Tang, Gennady Pekhimenko. Gist: Efficient Data Encoding for Deep Neural Network Training. *Proceedings of the 45th Annual Symposium on Computer Architecture (ISCA)*, June 2018.

[19] Jeremy Fowers, Kalin Ovtcharov, Michael Papamichael, Todd Massengill, Ming Liu, Daniel Lo, Shlomi Alkalay, Michael Haselman, Logan Adams, Mahdi Ghandi, Stephen Heil, Prerak Patel, Adam Sapek, Gabriel Weisz, Lisa Woods, Sitaram Lanka, Steven K. Reinhardt, Adrian M. Caulfield, Eric S. Chung, and Doug Burger. A Configurable Cloud-Scale DNN Processor for Real-Time AI. *Proceedings of the 45th Annual Symposium on Computer Architecture (ISCA)*, June 2018.

[20] N. P. Jouppi, C. Young, N. Patil, D. Patterson, G. Agrawal, R. Bajwa, S. Bates, S. Bhatia, N. Boden, A. Borchers, R. Boyle, P.-l. Cantin, C. Chao, C. Clark, J. Coriell, M. Daley, M. Dau, J. Dean, B. Gelb, T. V. Ghaemmaghami, R. Gottipati, W. Gulland, R. Hagmann, C. R. Ho, D. Hogberg, J. Hu, R. Hundt, D. Hurt, J. Ibarz, A. Jaffey, A. Jaworski, A. Kaplan, H. Khaitan, D. Killebrew, A. Koch, N. Kumar, S. Lacy, J. Laudon, J. Law, D. Le, C. Leary, Z. Liu, K. Lucke, A. Lundin, G. MacKean, A. Maggiore, M. Mahony, K. Miller, R. Nagarajan, R. Narayanaswami, R. Ni, K. Nix, T. Norrie, M. Omernick, N. Penukonda, A. Phelps, J. Ross, M. Ross, A. Salek, E. Samadiani, C. Severn, G. Sizikov, M. Snelham, J. Souter, D. Steinberg, A. Swing, M. Tan, G. Thorson, B. Tian, H. Toma, E. Tuttle, V. Vasudevan, R. Walter, W. Wang, E. Wilcox and D. H. Yoon, "In-Datacenter Performance Analysis of a Tensor Processing Unit," in *Proceedings of the 44th Annual International Symposium on Computer Architecture*, 2017.

[21] Swagath Venkataramani, Ashish Ranjan, Subarno Banerjee, Dipankar Das, Sasikanth Avancha, Ashok Jagannathan, Ajaya Durg, Dheemanth Nagaraj, Bharat Kaul, Pradeep Dubey, and Anand Raghunathan. 2017. ScaleDeep: A Scalable Compute Architecture for Learning and Evaluating Deep Networks. In *Proceedings of the 44th Annual International Symposium on Computer Architecture* (ISCA '17). ACM, New York, NY, USA, pp. 13-26.

[22] Angshuman Parashar, Minsoo Rhu, Anurag Mukkara, Antonio Puglielli, Rangharajan Venkatesan, Brueck Khailany, Joel Emer, Stephen W. Keckler, and William J. Dally. SCNN: An accelerator for compressed-sparse convolutional neural networks, *2017 ACM/IEEE 44th Annual International Symposium on Computer Architecture (ISCA)*, Toronto, ON, 2017, pp. 27-40.

[23] Yongming Shen, Michael Ferdman, and Peter Milder. 2017. Maximizing CNN Accelerator Efficiency Through Resource Partitioning. In *Proceedings of the 44th Annual International Symposium on Computer Architecture (ISCA)*. pp. 535-547

[24] Jiecao Yu, Andrew Lukefahr, David Palframan, Ganesh Dasika, Reetuparna Das, and Scott Mahlke. 2017. Scalpel: Customizing DNN Pruning to the Underlying Hardware Parallelism. *SIGARCH Comput. Archit. News* 45, 2 (June 2017), pp. 548-560.

[25] Christopher De Sa, Matthew Feldman, Christopher Ré, and Kunle Olukotun. 2017. Understanding and Optimizing Asynchronous Low-Precision Stochastic Gradient Descent. In *Proceedings of the 44th Annual International Symposium on Computer Architecture* (ISCA '17). ACM, New York, NY, USA, pp. 561-574.

[26] Jorge Albericio, Patrick Judd, Tayler Hetherington, Tor Aamodt, Natalie Enright Jerger, and Andreas Moshovos. 2016. Cnvlutin: ineffectual-neuron-free deep neural network computing. In *Proceedings of the 43rd International Symposium on Computer Architecture* (ISCA '16). IEEE Press, Piscataway, NJ, USA, pp. 1-13.

[27] Ali Shafiee, Anirban Nag, Naveen Muralimanohar, Rajeev Balasubramanian, John P. Strachan, miao Hu, R. Stanley Williams, and Vivek Srikumar. ISAAC: A Convolutional Neural Network Accelerator with In-Situ Analog Arithmetic in Crossbars, *2016 ACM/IEEE 43rd Annual International Symposium on Computer Architecture (ISCA)*, Seoul, 2016, pp. 14-26.

[28] Brandon Reagen, Paul Whatmough, Robert Adolf, Saketh Rama, Hyunkwang Lee, Sae Kyu Lee, José Miguel Hernández-Lobato, Gu-Yeon Wei, and David Brooks. 2016. Minerva: enabling low-power, highly-accurate deep neural network accelerators. In *Proceedings of the 43rd International Symposium on Computer Architecture* (ISCA '16). IEEE Press, Piscataway, NJ, USA, pp. 267-278.

[29] Kim Hazelwood, Sarah Bird, David Brooks, Soumith Chintala, Utku Diril, Dmytro Dzhulgakov, Mohamed Fawzy, Bill Jia, Yangqing Jia, Aditya Kalro, James Law, Kevin Lee, Jason Lu, Pieter Noordhuis, Misha Smelyankiy, Liang Xiong, and Xiaodong Wang. Applied Machine Learning at Facebook: A Datacenter Infrastructure Perspective. *2018 IEEE International Symposium on High Performance Computer Architecture (HPCA)*, Vienna, 2018, pp. 620-629.

[30] Yu-Hsin Chen, Joel Emer and Vivienne Sze, Eyeriss: A Spatial Architecture for Energy-Efficient Dataflow for Convolutional Neural Networks. *2016 ACM/IEEE 43rd Annual International Symposium on Computer Architecture (ISCA)*, Seoul, 2016, pp. 367-379.

[31] Denny Britz, Anna Goldie, Minh-Thang Luong, & Quoc Le. Massive Exploration of Neural Machine Translation Architectures. In *Proceedings of the 2017 Conference on Empirical Methods in Natural Language Processing,* pp. 1442-1451.

[32] TensorFlow. Accelerated Linear Algebra (XLA) Overview. https://www.tensorflow.org/performance/xla/, August 2018.

[33] ONNX: Open Neural Network Exchange Format. https://onnx.ai/, September 2018.

[34] Jason Power, Joel Hestness, Marc S. Orr, Mark D. Hill and David A. Wood, gem5-gpu: A Heterogeneous CPU-GPU Simulator, in *IEEE Computer Architecture Letters*, vol. 14, no. 1, pp. 34-36, 1 Jan.-June 2015.

[35] Rafael Ubal, Byunghyun Jang, Perhaad Mistry, Dana Schaa, and David Kaeli. Multi2Sim: a simulation framework for CPU-GPU computing. In *Proceedings of the 21st international conference on Parallel architectures and compilation techniques (PACT)*, pp. 335-34, 2012.

[36] AMD. MIOpen: AMD's Maching Intelligence Library. https://github.com/ROCmSoftwarePlatform/MIOpen, September 2018.

[37] AMD Radeon Technology Group. Radeon's next-generation Vega architecture. https://radeon.com/_downloads/vega-whitepaper-11.6.17.pdf, November 2017.

[38] Yifan Sun, Saoni Mukherjee, Trinayan Baruah, Shi Dong, Julian Gutierrez, Prannoy Mohan, and David Kaeli, "Evaluating Performance Tradeoffs on the Radeon Open Compute Platform," *2018 IEEE International Symposium on Performance Analysis of Systems and Software (ISPASS)*, 2018, pp. 209-218.

[39] Saiful A. Mojumder, Marcia S Louis, Yifan Sun, Amir Kavyan Ziabari, Jose L. Abellan, John Kim, David Kaeli, and Ajay Joshi. Profiling DNN Workloads on a Volta-based DGX-1 System. In 2018 *IEEE International Symposium on Workload Characterization (IISWC)*, 2018.

[40] Hongyu Zhu, Amar Phanishayee, Gennady Pekhimenko, Bianca Schroeder, Bojian Zheng, Mohamed Akrout, Andrew Pelegris, and Anand Jayarajan. Benchmarking and Analyzing Deep Neural Network Training. In 2018 *IEEE International Symposium on Workload Characterization (IISWC)*, 2018.

[41] AMD. Graphics Core Next Arcitecture, Generation 3. https://gpuopen.com/compute-product/amd-gcn3-isa-architecture-manual/. August, 2016.

[42] Minsoo Rhu, Mike O'Connor, Niladrish Chatterjee, Jeffrey Pool, Youngeun Kwon and Stephen W. Keckler. Compressing DMA Engine: Leveraging Activation Sparsity for Training Deep Neural Networks. In *2018 IEEE International Symposium on High Performance Computer Architecture (HPCA)*, Vienna, 2018, pp. 78-91.

[43] Angshuman Parashar, Minsoo Rhu, Anurag Mukkara, Antonio Puglielli, Rangharajan Venkatesan, Bruckey Khailany, Joel Emer, and Stephen Keckler. SCNN: An accelerator for compressed-sparse convolutional neural networks. In *2017 ACM/IEEE 44th Annual International Symposium on Computer Architecture (ISCA)*, Toronto, ON, 2017, pp. 27-40.

[44] Song Han, Xingyu Liu, Huizi Mao, Jing Pu, Ardavan Pedram, Mark A. Horowitz and William J. Dally. EIE: efficient inference engine on compressed deep neural network. In *2016 ACM/IEEE 43rd Annual International Symposium on Computer Architecture (ISCA),* Seoul, 2016, pp. 243-254

[45] Jorge Albericio, Patrick Judd, Alberto Delmas, Sayeh Sharify and Andreas Moshovos. Bit-pragmatic deep neural network computing. In *Proceedings of the 50th Annual IEEE/ACM International Symposium on Microarchitecture,* Boston, 2017, pp. 382-394.

[46] Patrick Judd, Jorge Albericio, Tayler Hetherington, Tor M. Aamodt, Natalie Enright Jerger, and Andreas Moshovos. Proteus: Exploiting Numerical Precision Variability in Deep Neural Networks. In *Proceedings of the 2016 International Conference on Supercomputing* (ICS '16).

[47] Tor Aamodt. COHESA: Computing Hardware for Emerging Intelligent Sensory Applications. https://www.ece.ubc.ca/~aamodt/projects/cohesa/.

[48] Luke Durant, Olivier Giroux, Mark Harris, and Nick Stam. Inside Volta: The World's Most Advance Data Center GPU. https://devblogs.nvidia.com/inside-volta/?ncid=so-lin-vt-13919, May 2017.

[49] AMD. AMD Graphics Core Next (GCN) Architecture. https://www.amd.com/Documents/GCN_Architecture_whitepaper.pdf , June 2012.

[50] Minjia Zhang, Samyam Rajbhandari, Wenhan Wang, and Yuxiong He. DeepCPU: Serving RNN-based Deep Learning Models 10x Faster. In *Proceedings of the 2018 USENIX Annual Technical Conference* (*USENIX ATC*), July 2018, pp. 951-965.

[51] John E. Stone, David Gohara, and Guochun Shi. 2010. OpenCL: A Parallel Programming Standard for Heterogeneous Computing Systems. *IEEE Design & Test* 12, 3 (May 2010), pp. 66-73.

[52] Matthew D. Sinclair, Johnathan Alsop, and Sarita V. Adve. 2017. Chasing Away RAts: Semantics and Evaluation for Relaxed Atomics on Heterogeneous Systems. In *Proceedings of the 44th Annual International Symposium on Computer Architecture* (ISCA '17). pp. 161-174.

[53] Matthew D. Sinclair, Johnathan Alsop, and Sarita V. Adve. 2015. Efficient GPU synchronization without scopes: saying no to complex consistency models. In *Proceedings of the 48th International Symposium on Microarchitecture* (MICRO-48). pp. 647-659.

[54] Yingying Tian, Sooraj Puthoor, Joseph L. Greathouse, Bradford M. Beckmann, and Daniel A. Jiménez. "Adaptive GPU cache bypassing." In *Proceedings of the 8th Workshop on General Purpose Processing using GPUs* (GPGPU). pp. 25-35.

[55] H.-H Lee, G. Tyson and M. Farrens. Eager writeback – a technique for improving bandwidth utilization. *In the proceedings of* the *33rd IEEE/ACM International Symposium on Microarchitecture (MICRO)*, 2000.

[56] J. Stuecheli, D. Kaseridis, D. Daly, H. C. Hunter and L. K. John. The virtual write queue: Coordinating DRAM and last-level cache policies. In the proceedings of *the 37th International Symposium on Computer Architecture (ISCA)*, 2010.

[57] C. Jeon, A. Li, L. Cox and S. Rixner. Reducing DRAM row activations with eager read/write clustering. *ACM Transactions on Architecture and Code Optimization (TACO)*, 10(4), p. 43, 2013.

[58] V. Seshadri, A. Bhowmick, O. Mutlu, P. B. Gibbons, M. A. Kozuch and T. C. Mowry. The dirty-block index. In the proceedings of *the 41st International Symposium on Computer Architecture (ISCA)*, 2014.

[59] Yifan Sun, Trinayan Baruah, Saiful A. Mojumder, Shi Dong, Xiang Gong, Shane Treadway, Yuhui Bao, Spencer Hance, Carter McCardwell, Vincent Zhao, Harrison Barclay, Amir Kavyan Ziabari, Zhongliang Chen, Rafael Ubal, José L. Abellán, John Kim, Ajay Joshi, and David Kaeli. "MGPUSim: Enabling Multi-GPU Performance Modeling and Optimization." In *Proceedings of 46th International Symposium on Computer Architecture* (ISCA), June 2019.

[60] Jonathan Lew, Deval Shah, Suchita Pati, Shaylin Cattell, Mengchi Zhang, Amruth Sandhupatla, Christopher Ng, Negar Goli, Matthew D. Sinclair, Timothy G. Rogers, and Tor M. Aamodt. Analyzing Machine Learning Workloads Using a Detailed GPU Simulator. In IEEE International Symposium on Performance Analysis of Systems and Software, ISPASS, 2019.

[61] Blake A. Hechtman, Shuai Che, Derek R. Hower, Yingying Tian, Bradford M. Beckmann, Mark D. Hill, Steven K. Reinhardt, and David A. Wood, "QuickRelease: A throughput-oriented approach to release consistency on GPUs," *2014 IEEE 20th International Symposium on High Performance Computer Architecture (HPCA)*, Orlando, FL, 2014, pp. 189-200.

[62] Singh, Inderpreet, Arrvindh Shriraman, Wilson WL Fung, Mike O'Connor, and Tor M. Aamodt. "Cache coherence for GPU architectures." In 2013 IEEE 19th International Symposium on High Performance Computer Architecture (HPCA), pp. 578-590. IEEE, 2013.

[63] Power, Jason, Arkaprava Basu, Junli Gu, Sooraj Puthoor, Bradford M. Beckmann, Mark D. Hill, Steven K. Reinhardt, and David A. Wood. "Heterogeneous system coherence for integrated CPU-GPU systems." In Proceedings of the 46th Annual IEEE/ACM International Symposium on Microarchitecture, pp. 457-467. ACM, 2013.

[64] Pei, Songwen, Myoung-Seo Kim, Jean-Luc Gaudiot, and Naixue Xiong. "Fusion coherence: scalable cache coherence for heterogeneous kilo-core system." In Advanced Computer Architecture, pp. 1-15. Springer, Berlin, Heidelberg, 2014.

[65] Orr, Marc S., Shuai Che, Ayse Yilmazer, Bradford M. Beckmann, Mark D. Hill, and David A. Wood. "Synchronization using remote-scope promotion." In ACM SIGPLAN Notices, vol. 50, no. 4, pp. 73-86. ACM, 2015.

[66] Alsop, Johnathan, Marc S. Orr, Bradford M. Beckmann, and David A. Wood. "Lazy release consistency for GPUs." In The 49th Annual IEEE/ACM International Symposium on Microarchitecture, p. 26. IEEE Press, 2016.

[67] Alsop, Johnathan, Matthew D. Sinclair, and Sarita V. Adve. "Spandex: a flexible interface for efficient heterogeneous coherence." In Proceedings of the 45th Annual International Symposium on Computer Architecture, pp. 261-274. IEEE Press, 2018.

[68] Vijaykumar, Nandita, Abhilasha Jain, Diptesh Majumdar, Kevin Hsieh, Gennady Pekhimenko, Eiman Ebrahimi, Nastaran Hajinazar, Phillip B. Gibbons, and Onur Mutlu. "A case for richer cross-layer abstractions: Bridging the semantic gap with expressive memory." In 2018 ACM/IEEE 45th Annual International Symposium on Computer Architecture (ISCA), pp. 207-220. IEEE, 2018.

[69] Li, Chao, Shuaiwen Leon Song, Hongwen Dai, Albert Sidelnik, Siva Kumar Sastry Hari, and Huiyang Zhou. "Locality-driven dynamic GPU cache bypassing." In Proceedings of the 29th ACM on International Conference on Supercomputing, pp. 67-77. ACM, 2015.

[70] Li, Ang, Gert-Jan van den Braak, Akash Kumar, and Henk Corporaal. "Adaptive and transparent cache bypassing for GPUs." In Proceedings of the International Conference for High Performance Computing, Networking, Storage and Analysis, p. 17. ACM, 2015.